\documentclass[prl,twocolumn,superscriptaddress,floatfix,showpacs]{revtex4-1}
\usepackage{graphicx}
\usepackage{dcolumn}
\usepackage{bm}
\usepackage{epsf}
\usepackage{subfigure}
\usepackage{amsmath}
\usepackage{amssymb}

\begin{document}

\title{Efficiency of isothermal molecular machines at maximum power}

\author{Christian Van den Broeck}
\affiliation{Hasselt University, B-3590 Diepenbeek, Belgium},
\author{Niraj Kumar}
\author{ Katja Lindenberg}
\affiliation{Department of Chemistry and Biochemistry and BioCircuits Institute, University
of California San Diego, 9500 Gilman Drive, La Jolla, CA 92093-0340, USA}


\begin{abstract}
We derive upper and lower bounds for the efficiency of an isothermal molecular machine operating at
maximum power. The upper bound is reached when the activated state is close to the fueling or
reactant state (Eyring-like), while the lower bound is reached when the activated state is
close to the product state (Kramers-like).
\end{abstract}

\maketitle

According to thermodynamics, different forms of work can be transformed into one another,
with an efficiency of at most $100\%$ \cite{callen}. This lossless limit is achieved with
a reversible process, i.e., an infinitely slow process. The corresponding power output
is therefore zero and thus of limited interest from a practical standpoint. 
One of the early discussions about efficiency at {\it finite power} is attributed to
Moritz von Jacobi around 1840. He realized that the output power of an
electrical device operating in the linear response regime is maximum when the
internal and external resistors are the same, yielding an efficiency of $50\%$. 
The Jacobi theorem can easily be reproduced in the much more general context of
{\it linear} irreversible thermodynamics: in any engine operating in the linear response regime,
maximum power is achieved when the loading force is equal to half of the stopping force;
the corresponding efficiency (output power over input power) is equal to $1/2$.  A
similar result has been proven for the transformation of heat into work,
where the maximum efficiency, the Carnot efficiency, is again achieved under reversible operation,
with zero power output. In the regime of
linear response the efficiency at maximum power is again $50\%$ of the
Carnot efficiency \cite{chris05}.  More recently, in this latter case 
various explicit results, including bounds for efficiency at maximum power, have also
been obtained in the {\it nonlinear}
regime \cite{ca751,ca752,ca753,ca754,ca755,ca756,ca757,esposito2009,esposito2010}. In
the present paper we show that similar results can be derived for {\it isothermal} molecular
machines \cite{howard,quarrie}.

\begin{figure}
 \centerline{\includegraphics[width=8cm]{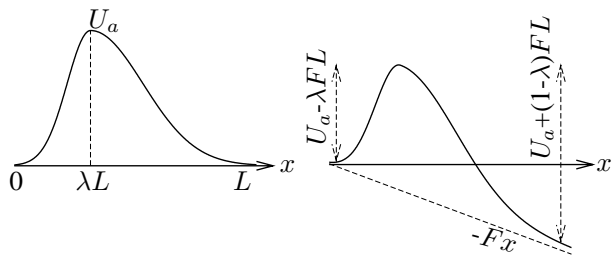}}
\caption{Schematic free energy potential $U_0(x)$ for a two-state molecular engine
described by a reaction coordinate $x$
under the net load force $F=F_1-F_2\ge 0$.}
\label{fig:mm}
\end{figure}

{\bf Generic model for a molecular motor}.
We first consider a generic model for a molecular motor, namely a two-state machine operating 
along a one-dimensional reaction coordinate, see Fig.~\ref{fig:mm}.
The states correspond to
two minima of an appropriate free energy landscape. While a physical energy landscape is expected
to be very complicated and  high-dimensional, the thermally activated transitions between the
two states will typically follow a preferred pathway that connects these states via the lowest
lying saddle point, the so-called activated state. One can project the motion on
this pathway and introduce a one-dimensional reaction coordinate $x$ with corresponding effective
free energy potential $U_0(x)$. The two ``rest" states of the  machine, that is, the minima in the
absence of external forces, correspond to, say, locations $x=0$ and $x=L$. The activated state lies
at an intermediate position $x_a=\lambda L$, $0\le \lambda \le 1$. In the unperturbed phase there are
no net transitions, and the states $1$ and $2$ have the same baseline potential value,
$U_0(0)=U_0(L)=0$. The potential has a
maximum $U_a=U_0(\lambda L)$ at the activated state, whose value is typically much larger
than the thermal energy $\beta^{-1}=k_B T$ ($T$ being the temperature and $k_B$ the
Boltzmann constant).
In this rest state, the rates, $k_0^+$ from $1$ to $2$ and $k_0^-$ from $2$ to $1$, are equal
and given by an Arrhenius law, $k_0^+=k_0^-\equiv k_0=\kappa \exp(-\beta U_a)$. We assume a constant
pre-exponential factor $\kappa$. 

In the operational regime, that is, in the presence of external forces, states $1$ and $2$ can
be identified as ``fuel" (or ``reactant") and ``product" states, respectively. To transform fuel
into product, the machine is subject to a driving force $F_1$ which allows it 
to overcome an opposing but weaker loading force $-F_2$, $F_2 \le F_1$. These forces
can be of various physical origins, including chemical (differences in chemical potentials),
electrical (internal or external electric fields) or mechanical (e.g., optical tweezers, atomic
force microscope or optical rotational torque). The combined effect of driving and loading
is a tilting of the potential towards the product state $2$, 
$U(x)=U_0(x)-Fx$, with $F=F_1-F_2\geq 0$. In a transition from state $1$ to state $2$,
a (scaled) input  energy $\epsilon_1=\beta F_1 L$ is transformed into a (scaled) output energy
$\epsilon_2=\beta F_2 L$. The efficiency of this transformation is given by
 \begin{equation}\label{efficiency}
\eta=\frac{\epsilon_2}{\epsilon_1}=\frac{F_2}{F_1}.
\end{equation}
Its maximum value, $\eta=1$, is reached  when the loading force $F_2$ approaches the
driving  force $F_1$, and  
the transition from $1$ to $2$ becomes infinitely slow.
In this reversible lossless limit the power vanishes.

In the case of finite and in particular of maximum power, the location of the
activated state plays a crucial role. For a so-called Eyring-like process \cite{howard}
the activated state is very close to the fuel state $1$, i.e., $\lambda$ is close to zero. 
The perturbation $-Fx$ barely affects the height of the activation barrier that needs to be
crossed to go from state $1$ to $2$. The rate also remains essentially unaffected,
$k^+\approx k_0$. However, a maximum barrier increase of $FL$ occurs for the backward
transition, resulting in a rate $k^-\approx k_0 \exp(-\beta FL)$ (assuming $FL \ll U_a$).
On the other hand, in the Kramers-like
scenario $\lambda \approx 1$ \cite{howard}, $k^+\approx k_0\exp(\beta F L)$, while $k^-\approx k_0 $
remains essentially unaffected. 
More generally, for a barrier at $x_a=\lambda L$, one has $k^+= k_0\exp(\lambda \epsilon)$
and $k^-= k_0 \exp[- (1-\lambda)\epsilon]$, where 
$\epsilon=\epsilon_1-\epsilon_2=\beta F L$ is the net energy loss or ``net load."
This is proper thermodynamic force (net force divided by the temperature) that appears in
the entropy production
and is thus a measure of the distance from equilibrium \cite{callen}.

With these explicit expressions for the rates, we turn to 
the output power $\Pi$ given by the output energy  $\epsilon_2$ multiplied by its net rate of
production, 
$\Pi=k\epsilon_2$, with $k=k^+ - k^-
= k(\epsilon)=k_0 \left[e^{\lambda \epsilon} -e^{-(1-\lambda) \epsilon}\right]$.
To specify the condition of maximum power we set
$\partial \Pi/ \partial \epsilon_2=0 $, which yields the unique solution
\begin{eqnarray}\label{estar}
\epsilon_2&=&\frac{1-e^{-\epsilon}}{\lambda(1-e^{-\epsilon})+e^{- \epsilon}}\\
&=&\epsilon+(\frac{1}{2}-\lambda)\,\epsilon^2+(\frac{1}{6}-\lambda+\lambda^2)\,\epsilon^3
+ O(\epsilon^4)\label{estar2}.
\end{eqnarray}
This result in Eq.~\eqref{efficiency} yields one of the central
results of this  paper, namely the efficiency at maximum power:
\begin{eqnarray}\label{main}
\eta^\star&=&\frac{e^\epsilon-1}{(\lambda \epsilon +1)(e^\epsilon-1)+\epsilon }\\
&=&\frac{1}{2}+\frac{1-2 \lambda}{8}\,\epsilon +\frac{1-12\lambda+12\lambda^2}{96}\,\epsilon^2+O(\epsilon^3).
\label{main2}
\end{eqnarray}

We point to a number of revealing observations. 
The first term of the expansion \eqref{main2} is the prediction of
linear irreversible thermodynamics of efficiency at maximum power equal to $1/2$.
The associated relation between the forces, $2 F_2=F_1$
is obtained from the first term in expansion
\eqref{estar2}, $\epsilon_2=\epsilon \equiv \epsilon_1 -\epsilon_2$.  

Turning to the next order corrections in Eqs.~\eqref{estar2} and \eqref{main2},
the coefficients vanish in the symmetric case $\lambda=1/2$, reminiscent of a
similar property for thermal machines \cite{esposito2009}. Note also that the coefficient
of the term proportional to $\epsilon$ in \eqref{main2}  goes from a maximum value
$1/8$  at $\lambda=0$ to the minimum value $-1/8$ for $\lambda=1$, switching from positive to
negative values at $\lambda=1/2$, again reminiscent of an analogous feature in thermal
machines \cite{esposito2010}.  The first two terms of the expansion \eqref{main2} were
also derived in \cite{seifert}, but the connection with the physically relevant parameter
$\lambda$ was not made (see, however, \cite{schmiedl}). 

The efficiency $\eta^\star$ at maximum power is a function of $\lambda$ and
$\epsilon$.  One
easily verifies that $\eta^\star$ is a monotonically decreasing function of $\lambda$
for given $\epsilon \ge 0$. The upper limit is the efficiency
$\eta^\star_E\equiv \eta^\star(0,\epsilon)\leq 1$
of the extreme Eyring-like scenario, and the lower limit is the efficiency
$0\leq \eta^\star(1,\epsilon)\equiv \eta^\star$ of the extreme Kramers-like case:
 \begin{equation}
\eta^\star_K \equiv  \frac{1-e^{-\epsilon}}{1-e^{-\epsilon}+\epsilon}\le \eta^\star \le
\frac{1-e^{-\epsilon}}{1-e^{-\epsilon}+\epsilon e^{-\epsilon}}\equiv \eta^\star_E.
\end{equation}

\begin{figure}
\centerline{\includegraphics[width=8cm]{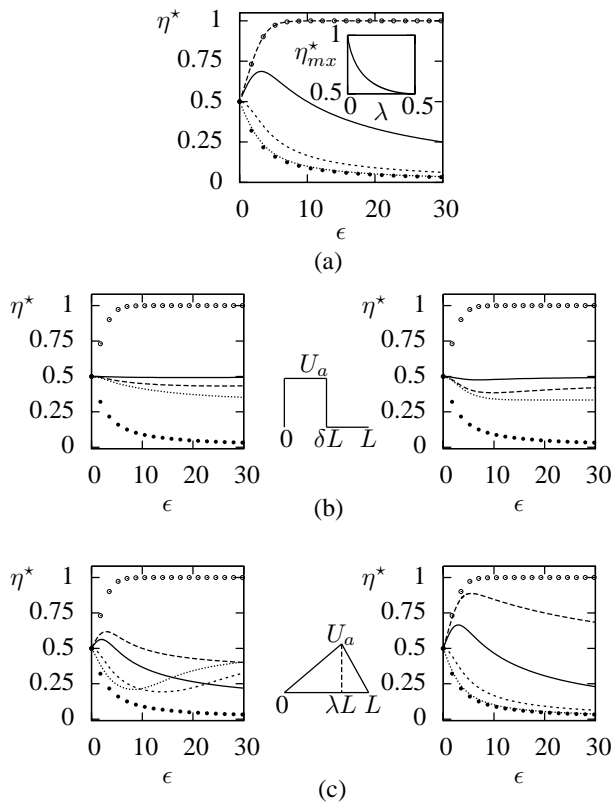}}
\caption{
Efficiency at maximum power, $\eta^{\star}$, as a function of net load $\epsilon$.
In all panels the open circles represent $\eta^*_E$ and the
filled circles $\eta^{\star}_K$.
Panel (a): two state model [cf. Eq.~(\ref{main})] with $\lambda=0$~(long dashes), $0.1$~(solid),
$0.5$~(short-dashes), and $0.9$~(dots).
The solid and long-dashed curves exhibit a maximum,
clearly seen on this scale in the
solid curve, while the two lower curves are monotonic. The inset shows the maximum value
of $\eta^*$ as a function of $\lambda$, with
$0 \le \lambda \le 1/2$.
The two panels in (b) are for the periodic square well potential,
and those in (c) are for the sawtooth potential.
For the former $\delta=0.1$ in the left panel, $0.5$ in the right panel. In both panels,
$\eta^\star$ is shown for three barrier heights:
$\beta U_a = 1$ (solid), $3$ (dashes), and $9$ (dots). For the sawtooth potential the
curves in the left panel are all for $\beta U_a=10$ and in the right panel for $\beta U_a = 100$.
The four curves are for different values of the potential asymmetry parameter $\lambda$:
$\lambda=0$ (long dashes), $0.1$ (solid), $0.5$ (short dashes), and $0.9$ (dots).}
\label{fig:estar}
\end{figure}

We next consider the $\epsilon$ dependence, starting with the variation of the bounds.
The Eyring-like efficiency $\eta^*_E$ increases monotonically from $1/2$ when $\epsilon \to 0$ 
to $\eta^*_E =1$ when $\epsilon \to \infty$.
The Kramers-like efficiency $\eta^*_K$ decreases monotonically from $1/2$ when $\epsilon \to 0$ to 
$\eta^*_K=0$ when $\epsilon\to \infty$.  The variation of $\eta^*$ between these bounds
depends on $\lambda$.  When $\lambda \geq 1/2$, $\eta^*$ decreases monotonically from
$1/2$ when $\epsilon =0$ to $0$ when $\epsilon\to\infty$. The system is in the
product regime (Kramers-like), and $\eta^*$ behaves much like
the Kramers-like limit, never rising above the linear response value $1/2$ [dotted curve,
short-dashed curve, and
filled circles in Fig.~\ref{fig:estar}(a)].  On
the other hand, when $0<\lambda\le 1/2$ the system is in the fuel regime (Eyring-like), and $\eta^*$
starts at $1/2$ when $\epsilon=0$, rises to a unique maximum, and then decreases to $0$ as $\epsilon
\to \infty$ [long-dashed curve, solid curve, and open circles
in Fig.~\ref{fig:estar}(a)].
The ``optimal" value of the efficiency at maximum power occurs at the 
net load value $\bar{\epsilon}$ which solves the transcendental equation obtained by
setting the derivative of \eqref{main} with respect to $\epsilon$ equal to zero,
$(1-e^{\bar{\epsilon}})\left[\lambda(1-e^{\bar{\epsilon}})-1\right] = \bar{\epsilon}
e^{\bar{\epsilon}}$.
Each point along the curve in the inset of Fig.~\ref{fig:estar}(a) is associated with a
different value of $\epsilon$. High efficiencies at maximum power require the system
to operate very near the fuel state.  Thus, for instance, referring to the figure, the
maximum of the $\lambda=0.1$ curve (solid) is $\eta^*=0.69$ and occurs when the net load is
$\epsilon=3.19$.  A maximum efficiency of say $\eta^*=0.9$ requires that the net load be
$\epsilon=5.68$ and that the motor operate at $\lambda=0.016$.  

We can repeat our analysis
for $\epsilon \le 0$, with net transitions going from state $2$ to state $1$. Indeed,
many motors, including ATPase, can operate in reverse.  The interchange of $1$ and $2$
corresponds to a replacement of $\lambda$ by $1-\lambda$. Hence the above theory indicates
that, when considering both modes of operation, at least one of them has $\lambda\ge 1/2$,
with a corresponding  efficiency at maximum power less than $50\%$. For $\lambda= 1/2$
the engine works equally well at maximum power in forward and reverse modes, at $50\%$ efficiency.

{\bf Generalized model}.
The above generic model assumes exponentially difficult
crossing of the transition state. With a more general analysis, we suppose
that the motion projected on reaction coordinate $x$ can be described
as an overdamped one-dimensional diffusion process in a potential $U(x)=U_0(x)-Fx$:
\begin{equation}{\label{e_s5}}
\gamma \frac{d x(t)}{dt}=-\frac{dU(x(t))}{dx}+\sqrt{2\gamma k_B T}\xi(t).
\end{equation}
Here  $\gamma$ is 
the viscous friction coefficient and $\xi$ is Gaussian white noise, 
$\langle \xi(t)\rangle=0$, and $\langle\xi(t)\xi(t^\prime)\rangle=\delta(t-t^\prime)$.
The diffusion coefficient is related to the viscous friction coefficient and the temperature
by the Einstein relation $D =k_B T/\gamma$.  This description 
relates transition rates to more fundamental parameters than the earlier phenomenology. 
It has been shown to be
a very useful tool to describe the response of molecular machines \cite{olga}, and it has
the further advantage of being  analytically tractable. It also does not require extremely high
activation barriers.  Indeed, when the potential barriers are
comparable to the thermal energy $k_B T$, it is no longer appropriate to identify the
minima of the potential (formerly $x=0$ and $x=L$) as states between which transitions
take place. We therefore replace the two-state scenario by an
extended coordinate $x \in [-\infty,\infty]$, and consider a baseline potential
$U_0(x)$ periodic in $x$ with period $L$.

Upon application of a driving force $F_1$ and a load force $F_2$, the net
force $F=F_1-F_2 \ge 0$ induces a steady state current in the tilted potential
$U(x)=U_0(x)-Fx$ with average velocity $V=\langle dx/dt \rangle$ along the
positive $x$-coordinate. The transformation of  driving energy
$\epsilon_1=\beta F_1 L $ per period into
loading energy $\epsilon_2=\beta F_2 L$  per period takes place at a net
rate $k=V/L$, with output power $\Pi=k \epsilon_2=V \epsilon_2/L$.
The efficiency of
the transformation is again given by $\epsilon_2/\epsilon_1$, see Eq.~\eqref{efficiency}. 
We note in passing that, besides being a natural model for cyclic molecular
motors such as ATPase, overdamped Brownian motion in a tilted periodic potential also
provides a relevant description in a large number of other physical situations~\cite{peter},
including Josephson junctions, rotating dipoles in external fields,
particle separation by electrophoresis, transport in tubes of varying cross-section,
and biophysical processes such as neural activity and intracellular transport.   
 
Turning to the issue of efficiency at maximum power, we first derive results that hold for
arbitrary potential. We suppose that the average velocity $V$ can be written as a power
series in $F$. Since $V$ vanishes for $F=0$, there is no constant term in the
expansion. For comparison with the previous results, we consider
the rate $k=V/L$ for moving over one period $L$, and write the corresponding power
series in terms of $\epsilon=\beta FL$,
$k=k(\epsilon)=V/L =a_1 \epsilon +a_2 \epsilon^2+a_3 \epsilon^3+O(\epsilon^4)$.
Maximization of the output power $\Pi=k(\epsilon) \epsilon_2 $ with respect to the
loading energy $\epsilon_2$
gives the following expansion for the output yield at maximum power [compare
with Eq.~\eqref{estar2}]:
\begin{equation}{\label{F_2}}
\epsilon_2 =\epsilon-\frac{a_2}{a_1} \epsilon^2+2(\frac{a_2^2}{a_1^2}-\frac{a_3}{a_1})
\epsilon^3+O\left(\epsilon^4\right).
\end{equation}
The corresponding efficiency reads [see Eq.~\eqref{main2}]:
\begin{eqnarray}{\label{effi-cubic}}
\eta^\star=\frac{1}{2}-\frac{a_2}{4 a_1}\epsilon+\frac{\left(3 a_2^2-4 a_1 a_3\right)}{8 a_1^2}\epsilon^2 
      +O\left(\epsilon^3\right).
\end{eqnarray}
This expansion features the familiar $50\%$ efficiency in the 
regime of linear response. 
Turning to the nonlinear regime, we note that just as in the generic two state model,
the next order correction vanishes ($a_2=0$)
whenever the system has left-right symmetry for the velocity, $V(F)=-V(-F)$.

The average steady state velocity
for overdamped motion in a tilted periodic potential
is given by~\cite{strat}
\begin{equation}{\label{e_n14}}
V(\epsilon)=\frac{D L \left(1-e^{-\epsilon }\right)}{\int_0^L\,dx \int_0^L \,dy\, e^{\beta\left[-U_0(x)+U_0(x+y)\right]-\epsilon y/L} }.
\end{equation}
This expression in principle allows us to find the power $\Pi=V \epsilon_2/L$, and hence
makes it possible to explicitly identify the regime of maximum power and its corresponding
efficiency. 

Firstly, we rederive the results for the two-state model with a potential $U_0$
with a dominant high maximum in each period, say at $x=x_a=\lambda L$ (modulo $L$), and a
unique minimum at $x=0$ (modulo $L$). 
The dominant contribution to the double integral in
Eq.~\eqref{e_n14} comes from
the region around the $(x,y)$ point for which $U_0(x+y)$ reaches a
maximum and $U_0(x)$ a minimum. This point lies at $x+y\equiv x_a=\lambda L$ 
and  $x=0$, and consequently $y=\lambda L$. The $\epsilon$-dependence of the
denominator is therefore of the form $\exp(-\lambda \epsilon)$.
It then follows directly that
 $V\sim k \sim \{1-\exp(-\epsilon)\} \exp(\lambda \epsilon)$, and the resulting
power is identical to that for the two-state model.

Secondly, we note that universal conclusions can be drawn using
Eq.~\eqref{e_n14} even without an
explicit evaluation of the integrals.  In particular, we can identify the coefficients $a_i$
in Eq.~\eqref{effi-cubic},
\begin{eqnarray}{\label{e_n19}}
 a_1&=&\frac{D}{L^2}\,\frac{1}{ I_0},\;\;\;\;\;\;a_2=\frac{D}{L^2}\,\frac{2I_1-I_0}{2  I_0^2},\nonumber\\
 a_3&=&\frac{D}{L^2}\,\frac{I_0^2+6I_1^2-3I_0\left(I_1+2I_2\right)}{6  I_0^3},
\end{eqnarray}
where 
\begin{equation}{\label{e_n17}}
I_n=\frac{1}{n!L^{n+2}}\int_{0}^{L}dx\, e^{-\beta U_0(x)}\int_{0}^{L}dy\, y^ne^{\beta U_0(x+y)}.
\end{equation}
Since $0 \le I_1\le I_0$,  the coefficient ${a_2}/{4 a_1}=(I_1/I_0-1/2)/4$ of the linear
term lies between $-1/8$ and $1/8$, as was the case
for the 2-state model, see Eq.~\eqref{main2}. Furthermore, the coefficient is
zero for a  potential with left-right symmetry, that is, when there exists a
point $x_0$ for which $U_0(x-x_0)=U_0(x_0-x)$.

To proceed further, it is in general necessary to invoke numerical calculations because
the integrals in \eqref{e_n14} can not be performed analytically for a general
potential $U_0(x)$.  However, analytic results can be obtained in
some limits (such as the dominant high maximum case considered above), or for some
specific shapes of the potential. Examples include the square well,
$U_0(x) = U_a,~x \in [0, \delta L]$ modulo $L$, and $U_0(x)
       = 0,~ x \in [\delta L, L]$ modulo $L$,
and a saw-tooth potential.
To illustrate, we  quote the  efficiency at maximum power for the square well potential,
\begin{eqnarray}\label{swp}
\eta^\star&=&\frac{(e^{\epsilon}-1)}
{(e^{\epsilon}-1)
(3 +\epsilon \zeta)+\epsilon}, \\
 \zeta&=& \frac{\epsilon(e^\epsilon-1) + c\left(1+e^\epsilon +e^{(1-\delta)\epsilon}
-e^{\delta\epsilon}\right)}{(1-\epsilon-e^\epsilon) +c\left[1-\delta
e^{(1-\delta)\epsilon}
+ (\delta-1)e^{\delta\epsilon}\right]},\nonumber
\end{eqnarray}
where $c\equiv 4\sinh^2(\beta U_\alpha /2)$.
The efficiency $\eta^\star$ for the square-well and sawtooth potentials are shown
in Figs.~\ref{fig:estar}(b) and (c).

Finally, we mention a generic behavior for any potential with finite maxima.
For very large driving, the barrier(s) of the potential $U_0$, indeed the entire potential,
become irrelevant and
one goes back to a linear model with $V=F/\gamma=a_1 \epsilon$ with $ a_1=D/L$,
implying that $\eta^{\star}=1/2$. This return to the linear scenario can
be seen in some cases in Fig.~\ref{fig:estar};
in others one must go to higher values of $\epsilon$ than
those shown in the figure.

{\bf Closing perspective}.
Is the issue discussed in this paper, maximizing
power with respect to the load, a relevant criterion in practice? In the case of
thermal motors, this seems to play a role, at least from an engineering point
of view, since power plants  operate under conditions in general agreement with this
criterion \cite{esposito2010}. We hope that the present paper will lead to a
re-examination, from the perspective of maximum power, of the much larger class
of isothermal engines, including the important class of molecular motors.

This work was partially supported by the US National Science Foundation under
Grant No. PHY-0855471.


\begin{thebibliography}{}
\bibitem{callen}
H. B. Callen, {\it Thermodynamics and an Introduction to
Thermostatistics} (Wiley, 2 ed., 1985).

\bibitem{chris05}
C. Van den Broeck,
Phys. Rev. Lett. {\bf 95}, 190602 (2005).

\bibitem{ca751} 
F. Curzon and B. Ahlborn,
Am. J. Phys. {\bf 43}, 22 (1975).

\bibitem{ca752}
T. Schmiedl and U. Seifert, 
EPL {\bf 81}, 20003 (2008).

\bibitem{ca753}
Z. C. Tu,
J. Phys. A {\bf 41}, 312003 (2008).

\bibitem{ca754}
A. E. Allahverdyan, R. S. Johal and G. Mahler, 
Phys. Rev. E {\bf 77}, 041118 (2008).

\bibitem{ca755}
Y. Izumida and K. Okuda,
EPL {\bf 83}, 60003 (2008).


\bibitem{ca756}
M. Esposito, K. Lindenberg and C. Van den Broeck, 
EPL {\bf 85}, 60010 (2009).

\bibitem{ca757}
Esposito, R.Kawai, K. Lindenberg and C. Van den Broeck, 
Phys. Rev. E {\bf 81}, 041106 (2010).

\bibitem{esposito2009}
M. Esposito, K. Lindenberg and C. Van den Broeck,
Phys. Rev. Lett. {\bf 102}, 130602 (2009).

\bibitem {esposito2010}
M. Esposito, R. Kawai, K. Lindenberg and C. Van den Broeck,
Phys. Rev. Lett. {\bf 105}, 150603 (2010).

\bibitem{howard}
J. Howard, {\it  Mechanics of motor proteins and the cytoskeleton}
(Sinauer Associates: Sunderland, MA, 2001).

\bibitem{quarrie}
D. A. McQuarrie, {\it Statistical Mechanics} (University
Science Books, Sausalito, CA,  1st ed., 2000). 

\bibitem{seifert}
U. Seifert,
Phys. Rev. Lett.  {\bf 106}, 020601 (2011).

\bibitem{schmiedl}
T. Schmiedl and U. Seifert, EPL {\bf 83}, 30005 (2008).

\bibitem{olga}
O.K. Dudko, T. G. W. Graham and R. B. Best, 
Phys. Rev. Lett. {\bf 107}, 208301 (2011).

\bibitem{peter}
P. Reimann, 
Phys. Rep. {\bf 361}, 57 (2002).

\bibitem{strat}
R. L. Stratonovich, {\it Topics in the Theory of Random Noise Vol. 1} (Gordon and Breach, New
York, 1963).

\end{thebibliography}
\end{document}